\documentclass{JHEP3}
\usepackage{amsmath}
\newcommand{\Logo}{ Hamevol \ }
\newcommand{\stt}{ \small\tt }
\newcommand{\sit}{ \small\it }

\title{{\tt Hamevol1.0}: a C++ code for differential equations based on Runge-Kutta algorithm. An application to matter enhanced neutrino oscillation}
\author{P. Aliani$^{a,b}$, V. Antonelli$^{b,c}$, M. Picariello$^{b,c}$, Emilio Torrente-Lujan$^{d,e}$\\
$^{a}$ Dept. de Physique, Universit\'e Libre de Bruxelles, Bruxelles, Belgium\\
$^{b}$ Dip. di Fisica, Universit\`a degli Studi di Milano, Milano, Italy\\ 
$^{c}$ I.N.F.N., Sezione di Milano, Milano, Italy\\
$^{d}$ Departamento de Fisica, Grupo de Fisica Teorica, Universidad de
Murcia, Murcia, Spain\\
$^{e}$ CERN-TH, CH-1211 Geneve 23, Switzerland}

\abstract{We present a C++ implementation of a fifth order semi-implicit 
Runge-Kutta algorithm for solving Ordinary Differential Equations.
This algorithm can be used for studying many different problems
 and in particular it can be applied for 
computing the evolution of any system whose Hamiltonian is known. 
We consider in particular the 
problem of calculating the neutrino oscillation probabilities in presence of matter interactions.
The time performance and the accuracy of this implementation is competitive with respect to the other analytical 
and numerical techniques used in literature.   
The algorithm design and the salient features of the code are presented and discussed and some explicit 
examples of code application are given.

\vspace{5cm}

\vfill

{e-mail addresses:
paliani@ulb.ac.be;\,~vito.antonelli@mi.infn.it;\,~marco.picariello@mi.infn.it;\\emilio.torrente-lujan@cern.ch\\

{\large \it Submitted to Comput. Phys. Commun.}}
}

\newpage

\begin{document}

\section{Program Summary}
\begin{itemize}
\item{{\it Title of the program}: Hamevol} 
\item{{\it Version number}: 1.0} 
\item{{\it Avaible at}: http://wwwteor.mi.infn.it/~antonell/programs/RKutta 
\, , and mirrors} 
\item{{\it Programming language}: C++} 
\item{{\it Platform}: any platform supporting a C++ compiler (examples: Linux,
Unix, Windows)} 
\item{{\it Tested on}:{Pentium PC, AMD PC}}
\item{{\it Memory requirements for execution}: Standard application: 40 Kbytes}
\item{{\it No. of bytes in distributed program, including test data, 
etc}: 235.000}
\item{{\it Keywords}: 
Numerical algorithms, Differential equations, 
Hamiltonian evolution, 
Oscillation,}
\item{{\it Nature of physical problem}: Numerical solution of Hamiltonian 
differential equations. Application to the numerical calculation of the 
oscillation probability for a quantum system (like, for instance, neutrinos of 
any kind propagating in a medium) }
\item{{\it Method of solution}: Algorithm based  on fifth order semi-implicit 
Runge-Kutta method}
\item{{\it Typical running time}: $\simeq$ 30 seconds for every single point 
on a Penthium IV PC}
\end{itemize}

\section{Introduction. The mathematical problem.}

The use of numerical algorithms and suitable computational techniques
has often been very useful to find the solution of difficult problems 
which are of interest for mathematics and other applied science.
This is particularly true in our days, when the relationships between 
information technology and other sciences are becoming closer and closer.

In this paper we discuss the adaptation of well known numerical techniques
to a general class of problems which are described by ordinary differential 
equations and we present some examples taken from physics.

The numerical code and algorithm we are presenting in this work is based on 
the implementation of the Runge-Kutta method and it can 
find significant applications in the study of different physical systems. 
In fact the evolution of every system can be fully described once we are able 
to solve the differential equations that drive this evolution.
The typical example is the case in which one wants to study a linear 
quantum system that is described by a vector 
$X \equiv (X_i, i=1..N)$, where $X_i$ are the elements of an appropriate basis
decribing the system.
The system of linear differential  equations we are interested in 
 can be put in the simple form:
\begin{equation}\label{schrodeq}
i \frac{d X}{d t} = H(t) X,
\end{equation}  
where $H$ is a matrix  determining the evolution of the system.

In the language of physics $H$ is the Hamiltonian of the system and 
Equation (\ref{schrodeq}) is the corresponding Schroedinger equation. 
It is clear, however, that 
the system of differential equations (\ref{schrodeq}) is very general and they 
can describe also problems of different nature in fields that are completely different from physics. Depending on the kind of problems one has to solve, 
the requirements which are fundamental for the solutions can be different.
This can suggest the choice of a particular algorithm in order to fulfill these
requirements in the best possible way.
For instance, as we are going to discuss later, in the physical problems we 
are intersted in, the delicate point is efficiency more than accuracy and this 
justifies the choiche of a particular version of adaptive Runge-Kutta method 
that, if properly adapted to our purpose, enable us to obtain satisfactory
results. 

 The solution of equation (\ref{schrodeq}) can be written in terms of the
fundamental system of solutions, or equivalently
  called evolution operator $U\left(t,t_0\right)$, 
defined by the expression:
\begin{equation}\label{evolutionOperator}
 U\left(t,t_{0}\right)=\exp{\left[- i H \left(t-t_{0}\right)\right]},
\end{equation}
where $t_0$ is the initial time at which we know the state of the system.
The simple formula given above is valid for a
time independent Hamiltonian. In the case in which the Hamiltonian of
the system is changing in the time (like, for instance, in presence of matter
effects), formula (\ref{evolutionOperator}) must be replaced by path-ordered 
exponential
\begin{equation}\label{evolutiontime}
 U\left(t,t_{0}\right)=P \exp{\left[- i \int_{t_{0}}^t d \tau H[\tau]\right]}.
\end{equation}

The code we are presenting here can be used to solve in an iterative way 
the system of differential equations appearing in Equation (\ref{schrodeq}). 
This is particularly important in the case of Hamiltonians which are 
explicitly time dependent or which contain terms fastly oscillating in time. 

The structure of this paper is the following.
We start discussing in section~\ref{fisica} some concrete examples, taken from
physics, of situations in which the application of the algorithm presented 
here is particularly suited. 
In the next section we present the algoritmic structure and salient aspects of 
the code we developed. This section includes a presentation of the algorithm, 
all the essential informations about the distribution, the main subroutines
and functions and about the way of working of a sample program.
In section \ref{finale} we draw our final conclusions.
An example of a sample program, with some specific Hamiltonian set up and the 
relative outputs are presented in the Appendices.

\section{The Physical motivations.}\label{fisica}

A very interesting example is given by the study of neutrino physics, which
has been one of the central topics of elementary particle physics in the last
years.
A detailed discussion about the main properties of neutrinos and the
relevance of their study for our knowledge of the intimate structure of
matter is beyond the scope of this work. Therefore we refer the interested
reader to the many reviews one can find in literature \cite{revneut}.
Here we just recall that during the last years the answer has been given
to the central question of neutrino physics, which puzzled physicists for
more than seventy years, that is to discriminate whether this particle is
massive or massless.

We know by now that neutrinos are massive and oscillating particles 
and the proof of this has been given by the important results obtained
mainly in the last decade (and especially in the very last years) by the
experiments looking for oscillation signals of neutrinos from different
sources: solar \cite{SNO,SK}, 
reactor \cite{KamLAND} and atmospheric \cite{atmospheric} 
neutrinos.\\
The great relevance of these results is confirmed by the fact that this is up
to now the strongest indication of oscillation we have in the leptonic sector
and it is impossible to accomodate it in the usual ``minimal version''
of the Standard Model (the theory describing very well the electroweak
interactions of elementary particles). One can say that neutrino oscillations 
is a hint for physical phenomena beyond to what is presently known.

All these experimental evidences in favors of the oscillation hypothesis
have proved that the flavor eigenstates of neutrinos, that is the
ones entering in the weak processes, are in fact quantum superspositions
of different mass eigenstates 
(at least three different mass
eigenstates are needed to explain the full set of experimental data, if one
doesn't take into account the controversial results of the LSND experiment).
During the evolution the composition of this quantum system can change giving
rise to the oscillation phenomenon detected in the experiments.

Any study of neutrino oscillation is necessarily based on the calculation
of the so called neutrino survival (or transition) probability. This is the
probability that a neutrino emitted by a source with a certain flavor, 
for instance an electronic neutrino emitted in the solar fusion processes,
remains with the same flavour (or is converted into a different flavor
neutrino) before reaching the detector.

Hence the basic quantity to compute is the survival probability in matter for a neutrino of a certain 
flavor: 
\begin{equation}\label{probability}
P\left(\nu_{i}\rightarrow\nu_{i};t,E_{\nu}\right)=
\left
| 
\langle \nu_{i}\left(t_0 \right)|U\left(t,t_{0}\right)|\nu_{i}\left(t_{0}\right)\right \rangle
|^2
\end{equation}
where $t_0$ is the inital time at which the neutrino is assumed to be
in the flavour eigenstate $\nu_i$ (with $i = e, \mu, \tau$) and $U\left(t,t_{0}\right)$ is the evolution operator given by Equation (\ref{evolutiontime}).

 In absence of matter the basis in which the Hamiltonian is diagonal
is simply the mass basis, whose eigenstates $\nu_{\alpha}$ are connected to the neutrino flavor eigenstates $\nu_i$ by means of the mixing matrix $U$: 
\begin{equation}\label{VObasis}
\nu_{i}=\sum_{\alpha} U_{i \alpha }\, \nu_{\alpha},\, \text{  $i=e,\mu,\tau$ and $\alpha=1,2,3$}\, .
\end{equation}

Using the same compact notation  of Equation (\ref{schrodeq}), we denote the set 
of the three neutrino mass eigenstates with the vector $\nu$, where 
$\nu \equiv (\nu_{\alpha}, \alpha = 1,2,3)$. This notation can be simply 
extended to the case in which one has more than three neutrinos \footnote{This is the case if one introduces also sterile neutrinos in the analysis}.
The Schroedinger equation describing the evolution of the system in vacuum under the relativistic approximation and in the hypothesis of equal momentum 
for different mass eigenstates is: 
\begin{equation}\label{schrodeq2}
i \frac{d \nu}{dt} = H^0 \nu
\end{equation}  
where
\begin{equation}\label{VOHamiltonian}
H^0 = Diag E_{\alpha} = \sqrt{p^2+m^2} \simeq E_{\nu}+
\left(\begin{array}{ccc}
m_{1}^{2}/2E_{\nu}&&\\
&m_{2}^{2}/2E_{\nu}&\\
&&m_{3}^{2}/2E_{\nu}\\
\end{array}\right)
\end{equation}

The last experimental data (both for atmospheric and solar neutrinos) have proved that to describe neutrino 
evolution one has to take into account also the modification of the oscillation pattern due to the very 
important effects of interaction with matter. This gives rise to the well known MSW effect \cite{MSW}.
The problem of calculating neutrino oscillation probability in presence of matter effects has been 
faced by many authors with different approaches, both numerical and analytical, in the case of two neutrino 
flavors~\cite{pet3,tov1,2neut}.
Exact solutions to the three neutrino MSW equations 
were derived \cite{Tor3neut, other3neut} for simple matter densities. 
Numerical algorithms for direct computation of the solar neutrino survival 
probability with all three active neutrinos have been presented in 
\cite{KIM,JSK,Kim:2000sm} .

The purpose of our code \Logo  is to calculate the electron-neutrino survival probability for a given neutrino energy, a given density profile, given neutrino masses and mixing matrix. This set of (three) oscillation probabilities will be used subsequently by other (Fortran) programs to calculate expected signals from diverse neutrino experiments. Due to the fact that both the mixing matrix and the density function are considered input parameters of \Logo, all kinds of neutrino oscillation problems may be tackled, including those relevant to anti-neutrinos only.

In presence of standard matter with arbitrary electron number density,
the propagation is usually well described by the following system of 
differential equations: 

\begin{equation}\label{inmatter}
i \frac{d \nu}{dt} = \left(H^0+\rho(t) U V U^\dagger \right)\, \nu , 
\end{equation}
where V is a matrix with $V_{11}$ as only non-zero element, 
$\rho(t)$, essentially a forward scattering amplitude, is proportional to the 
electron number density of the medium $N_e(t)$
\begin{equation}\label{densita}
\rho(t)=\pm\sqrt{2} G_F N_e(t) 
\end{equation}
and $U$ is the mixing matrix connecting the 
neutrino flavor eigenstates with the mass eigenstates.
In the case of three neutrino generations, adopting the 
Particle Data Group \cite{PDG} convention for the mixing matrix,
one gets:
\begin{equation}\label{Mixmatr}
\begin{array}{ccc}
\left(
\begin{array}{c}
\nu_{e}\\
\nu_{\mu}\\
\nu_{\tau}
\end{array}\right)
 & = &
\left(
           \begin{array}{ccc}
c_1 c_3 & s_1 c_3 & s_3 \\
-s_1 c_2-c_1 s_3 s_2 & c_1 c_2 - s_1 s_3 s_2 & c_3 s_2 \\
s_1 s_2 - c_1 s_3 c_2 & -c_1 s_2 -s_1 s_3 c_2 & c_3 c_2
           \end{array}
        \right) \times 
\left(
\begin{array}{c}
\nu_{1}\\
\nu_{2}\\
\nu_{3}
\end{array}\right)
\end{array}
\end{equation}
where $c_i\equiv\cos\theta_i$ and $s_i\equiv\sin\theta_i$ and the three 
mixing angles $\theta_1=\theta_{12}$, $\theta_2=\theta_{23}$, and 
$\theta_3=\theta_{13}$ roughly measure mixing between mass eigenstates 
(1-2), (2-3), and (1-3) respectively. 
We have neglected the CP violating phase, which is irrelevant in this problem.

The plus/minus sign in formula of Equation (\ref{densita}) is for 
neutrinos/antineutrinos respectively.
The time dependence of the electron density $N_{e}$ is a crucial factor 
involved in solving the evolution equation.

The difference of the eigenvalues of $H^0$ (
the inverse of the usually defined oscillation length)
is typically considered to be, in 
the solar case, of the
order
$$\frac{m_i^2-m_j^2}{2 E} 
\approx \frac{10^{-4}-10^{-5}\ eV^2}{ 1\ MeV}
\approx 10^{-10}-10^{-11}\
eV $$ 

In \cite{Tor3neut} an analytical solution of the problem has been found for the case in which the electron number density is parametrized (data taken from \cite{bah1}) , for sufficiently far distances from the solar core as:  
\begin{eqnarray}\label{paramtorr}
N_e(r)  = N_0 \exp(-\lambda r) & ; & \lambda \simeq 10.6 \, \frac{r}{r_0} 
\end{eqnarray}
with $r,r_0$ the distance from the center and solar radius respectively.

An important peculiarity of neutrino propagation in matter 
with respect to vacuum
is that for a certain value of the parameters
a resonance  appears at a certain point along the neutrino trajectory.
 This resonance takes place if the following condition is satisfied:
\begin{equation}\label{reseq}
\rho(t_{res}) = \frac{\Delta m^2 \cos 2 \theta}{E},  
\end{equation}
where $\theta$ is the mixing angle between the 2 neutrino flavors taking
part to the oscillation phenomenon. 
Apart from this approximate results, exact solutions have appeared for
particular forms of the function $\rho $:  for linear densities,
in terms of Weber-Hermite functions (\cite{pet2,hax1}); for functions
of the form $\rho(t)= C (1+\tanh (\lambda t)) $
in \cite{not1}, and for exponentially decaying densities
$\rho(t)= c e^{-\lambda t} $ in \cite{pet3,tov1}.

The parametrization of Equation (\ref{paramtorr}) is the one used also in the 
sample program we present as an example in subsection \ref{sample}.\\
In any case our numerical alghorithm enables us to find a solution of the problem for every expression one chooses for the electron density number.

The capability of solving the system of Equations (\ref{inmatter}) and, therefore, of computing the neutrino survival (or transition) probabilities as a function of the mixing parameters is an essential ingredient in every analysis of neutrino data. The theoretical expected signal in every experiment is obtained by convoluting neutrino fluxes, oscillation probabilities, neutrino cross sections and detector energy response functions. The comparison of these expected signal as a function of the mixing parameters with the experimental results is then usually performed by means of a $\chi^2$ statistical analysis.
The outcome of this kind of analyses is typically the production of exclusion plots selecting the regions of the mixing parameter plan which are in agreement with the data at a certain confidence level.
The algorithm we are presenting in this work enables us to numerically solve the neutrino evolution equations for all the oscillation parameter space, without the need to introduce the approximation which are required in different approaches based on the use of semi-analytical expressions in portions of the parameter space.
For a more detailed description of the full procedure we adopted for a phenomenological analysis of solar and reactor neutrino data we refer the interested reader to \cite{ourbasic}.

The numerical alghoritm we are presenting here finds also other relevant applications. 
Some significant examples are:\\
a)the study of the neutrino evolution inside stochastic media and neutrino propagation in solar magnetic field~\cite{Torsec}.
In these cases one has to solve systems of differential equations where, 
respectively, $(2 N)^2$ and $(2 N)$ equations appear (N is the number of neutrino species);\\
b) solution of the renormalization group equation for the Minimal Supersymmetric Standard Model in a Supergravity scenario. 
Here the number of differential equations in the systems that have
to be solved simultaneously (with a complicated mixture of initial and 
boundary conditions) is typically between thirty and one hundred.   
\cite{Torthird}
\section{Code Structure}
\subsection{Algorithmics}

There are different ways of solving a possibly stiff set of equations and 
the advantages and drawbacks of each of them must be evaluated keeping 
in mind the kind of problem one wants to study.
In our case we aim for efficiency rather than precision (a relative 
precision of $10^{-3}-10^{-5}$ in conversion probabilities for example would 
be sufficient in a typical application in particle physics propagation problems). Therefore we opted for an adaptive Runge-Kutta (RK) algorithm. 

The Runge-Kutta method \cite{NumericalRecipes} is particularly suitable for solving differential equations, starting from the knowledge of the function at the 
a fixed intial point $X$ and advancing the solution from $X$ to $X+h$, by  using the evaluation of the function at intermediate points inside the interval $h$
.
By properly combining these evaluations one can reduce the error in the final 
output. The method is conventionally denoted of order n if its error term is 
$O \left(h^{n+1}\right)$  

An essential characteristic of a good Ordinary Differential Equations integrator is the capability of having an adaptive control over its progress and a mechanism for adapting its stepsize, in such a way to obtain the required accuracy with
the minimal possible computational effort. This property is possessed by the 
algorithm we implemented.
Although an implicit RK would be advised for stiff equations, there are several alternatives, such as the semi-implicit fifth-order RK routine we
have chosen. 
This routine requires the determination of the function at five different points in the interval between the chosen steps. They are
\begin{eqnarray}\label{points}
k_{1}&=&hf\left(x_{n},y_{n}\right)\nonumber\\
k_{2}&=&hf\left(x_{n}+a_{2}h,y_{n}+b_{21}k_{1}\right)\nonumber\\
&&\ldots\\
k_{6}&=&hf\left(x_{n}+a_{6}h,y_{n}+b_{61}k_{1}+\ldots b_{65}k_{5}\right)\nonumber\\
y_{n+1}&=&y_{n}+\sum_{i=1}^{6}c_{i}k_{i}\nonumber
\end{eqnarray}
where the coefficients $a_{i}$, $b_{ij}$ and $c_{i}$ must satisfy certain
constraints in order to ensure stability and convergence. This algorithm is
well suited for adaptive stepping, due to the fact that among the six
evaluations in Eq. (\ref{points}) there is an embedded fourth-order
combination, which, although redundant, gives us an estimate of the error at
each evaluation thereby allowing us to adjust the step size. Table (\ref{CKcoeff}) gives a list of $a_{i}$, $b_{ij}$ and
$c_{i}$ as determined by Cash and Karp 
\cite{NumericalRecipes}.
\begin{table}[h]
\begin{center}

\vskip 0.2truecm
\begin{tabular}{lcccccccc}
\hline\hline
$i$&$a_{i}$&\multicolumn{5}{c}{$b_{ij}$}&$c_{i}$&$c_{i}^{*}$\\
\hline\\
\vspace{5pt}
1&              &              &               &             &&&$\frac{37}{378}$ &$\frac{2825}{27648}$\\
\vspace{5pt}
2&$\frac{1}{5}$ &$\frac{1}{5}$ &               &             &&&$0$              &$0$\\
\vspace{5pt}
3&$\frac{3}{10}$&$\frac{3}{40}$&$\frac{9}{40}$ &             &&&$\frac{250}{621}$&$\frac{18575}{48384}$\\
\vspace{5pt}
4&$\frac{3}{5}$ &$\frac{3}{10}$&$-\frac{9}{10}$&$\frac{6}{5}$&&&$\frac{125}{594}$&$\frac{13525}{55296}$\\
\vspace{5pt}
5&$1$           &$-\frac{11}{54}$&$\frac{5}{2}$&$-\frac{70}{27}$&$\frac{35}{27}$&&$0$&$\frac{277}{14336}$\\
6&\vspace{5pt}
$\frac{7}{8}$&$\frac{1631}{55296}$&$\frac{175}{512}$&$\frac{575}{13824}$&$\frac{44275}{110592}$&$\frac{253}{4096}$&$\frac{512}{1771}$&$\frac{1}{4}$\\
\hline
\multicolumn{2}{c}{$j=$}&1&2&3&4&5&\\
\hline\hline
\end{tabular}
\end{center}
\caption{Cash-Karp coefficients for our RK routines, taken from Ref.\protect\cite{NumericalRecipes}.}
\label{CKcoeff}
\end{table}

\subsection{The distribution}
The distribution of the program is contained in the 
tarred gzipped file
{\stt hamevol1.0.tar.gz}.
In any Linux or Unix system, unpacking and untarring this distribution 
file will produce a local directory called {\stt Hamevol1.0}
containing the following ascii files:

\begin{tabular}{lll}
&  &\\
&{\stt hamevol-rungekutta.hpp} &  The file with the main routines.  \\
&{\stt hamevol-util.hpp} &  Some auxiliary utilities.  \\
&{\stt hamevol-sample.hpp} & The sample program header.\\
&{\stt hamevol-sample.cpp} & The sample program.\\
&{\stt Makefile} & A simple compiling make example.\\
&  &\\
\end{tabular}

In the same Linux or Unix system the execution of the command

\begin{tabular}{lll}
&  &\\
&{\tt ./make } & \\
&  &\\
\end{tabular}

or directly the explicit call to the C++ compiler

\begin{tabular}{lll}
&  &\\
&{\tt g++ -O3 -o hamevol.x hamevol-sample.cpp} & \\
&  &\\
\end{tabular}

should produce an executable file from our sample 
program dedicated to 
solve some particular neutrino propagation problem.
To run the executable the following command should be typed 

\begin{tabular}{lll}
&  &\\
&{\tt ./hamevol.x {\it OPTION}}& \\ 
&  &\\
\end{tabular}

where in our sample program {\sit OPTION=1,0} depending on whether 
full Sun+Earth or only Sun propagation is demanded. As a consequence, 
some brief ouput will appear on the standard ouput, mainly information about parameter settings and 
options. In addition,
as a result of the execution our sample program  writes an 
output file {\stt runge.out} with the neutrino conversion 
probabilities along the neutrino trajectory.

In our distribution, we have well separated the code corresponding to
the general routines implementing the Runge-Kutta algorithm from those 
corresponding to a particular application (the ``sample'' files) 
of interest to us and that are presented here: the computation of oscillation 
neutrino probabilities. Within our sample programs, it is also well differentiated 
the driver code which calls the RK routines from the part where a concrete 
 hamiltonian is built. In the most basic case, 
a general user should be able to use our program as a black box for his 
own purposes simply plugging his own definition for the hamiltonian.

In the following we will first describe the main routines included in file 
{\stt hamevol-rungekutta.hpp} and then those contained in the sample programs.

\subsection{The RK algorithm. Main Subroutines}

The kern code is built up with five main subroutines. 
They correspond to procedures and methods well known in the literature.
We have improved and adapted them for our purposes.
The following classes are located in file {\stt hamevol-rungekutta.hpp}.
Here it follows a brief description of any of them together with its 
calling sequence. For brevity, arguments which are also referenced somewhere 
else are omitted here.

\begin{itemize}

\item 
{\small\tt 
void runge(CNumber y[], CNumber dydx[],  CNumber (*H)(...),
           int n, Number x, \\
Number h, CNumber yout[], void (*derivs)(...))
}

Given the value {\small\tt y[1,..,n]} of the vector state describing a
 phyisical system made up with {\stt n} components and evolving 
according to 
Schroedinger equation and knowing the Hamiltonian {\stt H} of the system, 
the subroutine produces the advanced solution as the function at the 
incremented variables {\small\tt yout[1..n]}.

\item
{\small\tt
void odeint(CNumber ystart[], int nvar, Number x1, Number x2, Number eps, \\
            Number h1, Number hmin,
            CNumber (*H)(...),
            void (*derivs)(...),
            void (*rkqs)(...))}

This is a Runge-Kutta driver with adaptive stepsize control. 
It Integrates the 
starting values {\small\tt ystart[1..nvar]} from 
{\stt x1} to {\stt x2} with accuracy {\stt eps}, storing the 
intermediate results in global variables. 

A value {\small\tt h1} should be set as a guessed first stepsize,
{\small\tt hmin} as the minimum  allowed stepsize (it can be zero). 
On output {\small\tt nok} and {\small\tt nbad} are the  number of good 
and bad  (but retried and fixed) steps taken, 
and {\small\tt ystart} is replaced by values at 
the end of the integration interval.

\item
{\small\tt
void rkqs(CNumber y[], CNumber dydx[], int n, Number *x, Number htry,
          Number eps, \\
CNumber yscal[], Number *hdid, Number *hnext,
          CNumber (*H)(...),
          void (*derivs)(...))}.
                  
This routine {\small\tt rkqs} is implemented in order to perform an 
adaptive 5th order Runge-Kutta integration. The method enables to have
 a monitoring of local truncation error, in order to ensure the 
required accuracy and adjust the stepsize. The inputs are the 
independent variable vector {\small\tt y[1..n]} and 
its derivative {\small\tt dydx[1..n]} at the starting value of the 
independent variable x. 
Other inputs are the stepsize {\small\tt htry}, 
the required accuracy {\small\tt eps}, 
and the vector {\small\tt yscal[1..n]} against which the error 
is scaled. 
On output, 
{\stt y} and {\stt x} are replaced by their new values, 
{\small\tt hdid} is the stepsize that was actually accomplished, 
and {\small\tt hnext} is the estimated next stepsize.

\item {\small\tt 
void rkck(CNumber y[], CNumber dydx[], int n, Number x, Number h,
          CNumber yout[], \\
CNumber yerr[],
          CNumber (*H)(...),
          void (*derivs)(...))}

 Used in adaptive size Runge-Kutta integration. Given values 
for the variables 
 {\stt y[1..n]} and their derivatives {\stt dydx[1..n]}
 known at {\stt x}, advance solution over an  interval {\stt h} and 
return the incremented 
variables as 
{\stt yout[1..n]}. 
Also returns an estimate of the local truncation error in yout 
using embedded fourth-order  method. 

The user supplies the routine {\stt H(x,i,j)}, which returns the element
{\stt (i,j)} of the hamiltonian of the evolution.

\item 
{\small\tt 
void deriv(Number x, CNumber y[], CNumber dy[], int n,
           CNumber (*H)(...))}.

The user-supplied routine {\small\tt derivs} is used for calculating 
the right-hand side derivative.
The user supplies the routine {\small\tt derivs(x,y,dydx,n,H)}, which returns the derivatives 
{\small\tt dydx} of the many variable function {\stt y}
 with respect to the vector {\stt x} at the point {\stt x}.

\end{itemize}

The following Auxiliary functions which will allocate 
 the following data structures are 
included in the auxiliary distribution file {\stt hamevol-util.hpp}.

\begin{tabular}{lll}
& & \\
& {\small\tt CNumber *Cvector(long nh)}: & 
allocate a {\stt CNumber} vector with subscript range {\small\tt v[1..nh]}. \\
&{\small\tt Number *vector(long nh)}:&  
      a  vector with subscript 
range {\small\tt v[1..nh]}.\\
& {\small\tt int *ivector(long nh)}:&
a vector with subscript range {\small\tt v[1..nh]}. \\
&{\small\tt unsigned long *lvector(long nh)}:&
a vector with subscript range {\small\tt v[1..nh]}.\\ 
&{\small\tt unsigned char *cvector(long nh)}:&
a vector with subscript range {\small\tt v[1..nh]}.\\
&{\small\tt template <class Type> Type **matrix(...)} &
allocate a {\stt Type matrix} with a subscript 
range. \\
& & \\
\end{tabular}

\subsection{Sample Program and Inputs}\label{sample}

The {\stt void main} routine in  {\stt hamevol-sample.cpp} file
takes the values of the neutrino wave functions at initial 
starting points for physical initial conditions which are standard 
for solar neutrino physics ($\nu_e(0)=1,\nu_\mu(0)=0,\nu_\tau(0)=0$) 
and calculates the final wave function and corresponding 
probabilities at the target final points. The algorithm includes 
the following steps:

\begin{itemize}
\item 
It takes from the command line the user argument ``1'' (Sun) or
``0'' (Earth) propagation.

\item It performs argument validation, set internal flags and 
writes to the standard output a list of current values of diverse
 parameters.

\item It declares the output file stream {\stt out ``runge.dat''},
 class {\stt ofstream} included within {\stt <fstream.h>}.

\item It declares the {\stt Cvector} objects {\stt nu,dnu}, respectively
 instances of the neutrino wave function and their vector derivative.
 It performs diverse other initializations.

\item Finally functions {\stt odeint} and {\stt evolve} are 
  repeteadly called until the desired final point or the 
 maximum number of steps  is reached. 

\end{itemize}

Different parameters, for example the number of equations, or 
 in this physical case the number of neutrino species ($N=2,3$), 
have to be set in the header file {\stt hamevol-sample.hpp} .
We 
run the RK algorithm to obtain the transition probabilites 
of neutrinos produced at the sun center with a given 
energy and mixing angle as a function to its position along 
the trajectory sun-earth.
The user should provide routines for computing the 
electron densities at Sun and the Earth along the 
neutrino trajectory. They appear in the matter part 
of the neutrino hamiltonian. We include examples of the main 
program and other smaller routines as appendices.

\section{Conclusions}\label{finale}

A new code based on a semi-implicit fifth order adaptive Runge-Kutta algorithm has been developed by us.
It can be used as solver for many systems of differential equations, like, for 
instance, the ones that usually describe the evolution of a system in physics 
and in other fields.
This algorith is particularly suited for the solution of differential equations
in which the operator driving the evolution of the system is changing in time.
 
Here we focus our attention to the application of this code to the study
of physical problems, like solving the Schroedinger equation for a system that 
is a quantum superposition of different possible states.
The explicit example we present is the study of the evolution and calculation 
of transition probabilty for neutrinos emitted by a source and travelling in 
a medium. This code has been already applied by us as a useful tool
to obtain a check with respect to other possible numerical algorithms (like 
the ones based on the evolution operator formalism) in our  
phenomenological analysis of different neutrino oscillation experiments.
This analysis has confirmed the validity of neutrino oscillation hypothesys 
and enabled us to determine the allowed region for mixing parameters, a topic 
of great relevance in Elementary Particle Physics.

In this paper we discuss the structure of the algorithm we developed and the 
main features of our code. We also present a sample program and give some 
typical outputs, as a concrete example of application of our algorithm.

\section{Acknowledgments}

We are really grateful to R. Ferrari for many useful discussions and 
suggestions and for the continuos support given to our work.
One of us (V. A.) would like to thank S. Petcov for very useful discussions 
about 
the analytical study of the neutrino progragation in matter.
One of us (E.T) would like to thank the hospitality of CERN-TH
division, the department of physics of University of Milan, and
financial support of INFN-CICYT grant. 
The computations presented here have been all performed on the computer 
farm of the Theoretical group of Milano University.


\newpage

\section{Appendices}

\subsection{Using Hamevol}

Here we present the main sample program, corresponding to the file {\stt hamevol-sample.cpp}, where we show explicitly the use of the main routines.

{\small
\begin{verbatim}

  #include "hamevol.hpp"
  #include <string.h>
  
  /* Inizialize the vacuum values  (.......)*/
  /* #define parameters            (.......)*/
  

  Number Var;
  CNumber **HH0;   // is the vacuum Hamiltonian in the flavour eingenstates
  struct mix{      // contains the mixing matrix
    Number mass[N + 1];
    CNumber U[N + 1][N + 1];
  } mixing;
  
  /* nu are the wave functions, dnu their derivative */
  CNumber *nu, *dnu;
  
  int main(int arg, char** argv){
    Number x1=0.;
    Number x2;
    
  /*  Argument and parameter validation (............) */
  /*   Information output               (.............)*/
    
    time_t ti, tf;
    CNumber *onu;
    
    nu = Cvector (N);
    dnu = Cvector (N);
    onu = Cvector (N);
    
    ofstream out ("runge.dat");	/* save data in runge.dat */
    
    srand (time (&ti));
    time (&ti);
    
    Var = VarI;
    Number VarO = Var;
    Number h = (VarF - VarI) / INIT_STEPS;
    Number eps = Eps_Error;
    vacuum_values ();
    
    /* Main Routines */
    odeint (nu, N, x1, x2, eps, dist, dist_min, H, deriv, rkqs);
    evolute (nu, &out, VarO);
    
    for (int nstp = 1; nstp <= MAX_STEPS; nstp++)
    {				/* Take at most MAXSTP steps */
      for (int i = 1; i <= N; i++)
      onu[i] = nu[i];
      vacuum_values ();
      if ((VarO + h - VarF) * (VarF - VarI) > 0.0)	/* Are we done? */
      h = VarF - VarO;
      Var = VarO + h;
      odeint (nu, N, x1, x2, eps, dist, 0.0, H, deriv, rkqs);
      if (distance (nu, onu, N) > Prob_Error)
      {
	//cout << "DECREASE: h=" << h << endl;
	h *= DECREASE;
	Number htemp = ((h < 0) ?
	FMIN (h, -abs ((VarI - VarF) / MAX_STEPS)) :
	FMAX (h, abs ((VarI - VarF) / MAX_STEPS)));
	if (htemp != h)
	{
	  evolute (nu, &out, Var);
	  VarO = Var;
	}
	h = htemp;
      }
      else
      {
	evolute (nu, &out, Var);
	h *= INCREASE;
	Number htemp = ((h < 0) ?
	FMAX (h, -abs ((VarI - VarF) / MIN_STEPS)) :
	FMIN (h, abs ((VarI - VarF) / MIN_STEPS)));
	if (htemp != h)
	h = htemp;
	VarO = Var;
      }
      
      if ((Var - VarF) * (VarF - VarI) > 0.0)
      {			/* Are we done? */
	cout << "t=" << time (&tf) - ti << endl;
	return 0;		/* normal exit */
      }
    }
    cout << "Too many steps in routine evolution_matter!" << endl;
    return -1;
}


\end{verbatim}
}

\subsection{Example of Hamiltonian definition}

Here is the hamiltonian we use in our sample program:

{\small
\begin{verbatim}

/************************************************************************/
/*                        The Hamiltonian                               */
/************************************************************************/


inline CNumber V(int i, int j){
  return (i == j == 1) ? 1. : 0.;
}
inline CNumber U(int i, int j){
  return mixing.U[i][j];
}

CNumber H(Number r, int i, int j){
  CNumber HH = 0;
  return HH0[i][j] + V (i, j) * rho (r) * sqrt (2.) * Gf;
}

void vacuum_values(){
  mixing.mass[1] = 1.e-2;
  mixing.mass[2] = 1.e-1;
  Number th12 = M_PI / 3.;
  Number th13 = M_PI / 3.;
  Number th23 = M_PI / 3.;
  
  nu[1] = 1.;
  nu[2] = 0.;

  CNumber H0[N + 1][N + 1];

  HH0 = matrix ((CNumber) 1, N, 1, N);
  Number sth12 = sin (th12);
  Number cth12 = cos (th12);
  Number sth13 = sin (th13);
  Number cth13 = cos (th13);
  Number sth23 = sin (th23);
  Number cth23 = cos (th23);
  
   /*   Three neutrinos   */

  (mixing.U)[1][1] = CNumber (cth12 * cth13);
  (mixing.U)[1][2] = CNumber (sth12 * cth13);
  (mixing.U)[1][3] = CNumber (sth13);
  (mixing.U)[2][1] = CNumber (-sth12 * cth23 - cth12 * sth23 * sth13);
  (mixing.U)[2][2] = CNumber (cth12 * cth23 - sth12 * sth23 * sth13);
      (mixing.U)[2][3] = CNumber (sth23 * cth13);
      (mixing.U)[3][1] = CNumber (sth12 * sth23 - cth12 * cth23 * sth13);
      (mixing.U)[3][2] = CNumber (-cth12 * sth23 - sth12 * cth23 * sth13);
      (mixing.U)[3][3] = CNumber (cth23 * cth13);
      break;
    default:
      cerr << "Number of neutrina (" << N << ") not implemented!" << endl;
      exit (-1);
    }
  /*  
     cout << "Vacuum values:\n";
     cout << "- Hamiltonian:\n";
     Hamiltonian HH = H0();
     cout << HH;
     cout << "- Mixing matrix:\n";
     cout << *(mixing.U);
     */
     
     for (int i = 1; i <= N; i++)
     for (int j = 1; j <= N; j++)
     {
       HH0[i][j] = 0;
       /* H0 is the vacuum Hamiltonian in the mass eingenstates */
       if ((i == 1) && (j == 1))
       H0[i][i] = 1. / pow (10, Var);	// This is OK for two neutrina
       else
       H0[i][j] = 0;
       /* HH0 is the vacuum Hamiltonian in the flavour eingenstates */
       for (int k = 1; k < N; k++)
       for (int l = 1; l < N; l++)
       HH0[i][j] += conj (U (k, i)) * H0[k][l] * U (l, j);
     }
     
     
     return;
}

\end{verbatim}
}

The user should provide routines for computing the 
electron densities at Sun and the Earth along the 
neutrino trajectory. They appear in the matter part 
of the neutrino hamiltonian.

\subsection{Sample outputs}

The following is the verbatim output of our program 

\noindent
{\stt hamevol.x 0} 

for the  values
of the parameters which appears in the first information lines (included
 by default in the sample program).

{\small
\begin{verbatim}


./hamevol.x 0
Starting evolution in the Sun
Used parameters:
MAX_STEPS   100000      INIT_STEPS  10000
DECREASE    0.1 INCREASE    5
VarI        -2.39794    VarF        -12.3979
Eps_Error   1e-08       Prob_Error  0.01
x2/VarI     0.00881916  x2/VarF     8.81916e+07

  -2.3979       1	0.00382	4.62e-44	  4e-08
  -2.4979       1	0.00382	4.62e-44	  4e-08
  -2.5979       1	0.00481	4.62e-44	-1.4e-08
  -2.6979       1	0.00605	4.62e-44	-8.5e-08
  -2.7979       1	0.00762	4.62e-44	1.5e-09
  -2.8979       1	0.00959	4.62e-44	-1.7e-08
  -2.9979       1	 0.0121	4.62e-44	3.6e-08
  -3.0979       1	 0.0152	4.62e-44	-2.8e-08
  -3.1979       1	 0.0191	4.62e-44	7.8e-08
    
     (......................)
   
\end{verbatim}
}

\begin{table}[p]
\centering
\begin{tabular}{|l|c|l|}
 \hline\vspace{0.1cm}
Variable & default & description\\[0.1cm]
  \hline\\ 
{\stt hamevol-rungekutta.hpp} & &\\ 
{\small\tt MAXSTP}    & 1000000               & RK algorithm internal parameter\\
{\small\tt TINY}      & $1.0 \times 10^{-10}$ & id. \\
{\small\tt SAFETY}    & 0.9                   & id. \\
{\small\tt PGROW}     & -0.2                  & id. \\
{\small\tt PSHRNK}    & -0.25                 & id. \\
{\small\tt ERRCON}    & $1.89 \times 10^{-4}$ & id. \\
{\small\tt }          &                       & \\
{\stt hamevol-sample.cpp} & &\\ 
{\small\tt MAX-STEPS} & $1.0 \times 10^{5}$   & Steeper method\\
{\small\tt MIN-STEPS} & 10000                 & id.\\
{\small\tt INIT-STEPS}& 10000                 & id.\\
{\small\tt DECREASE}  & 0.1                   & id.\\
{\small\tt INCREASE}  & 5.0                   & id.\\
{\small\tt }          &                       & id.\\
{\stt hamevol-sample.hpp} & &\\ 
{\small\tt fermi-MeV} &1.0/197.326            & conversion  $f\to 1/MeV$\\ 
{\small\tt m-eV}& fermi-MeV 
$\times \frac{10^{15}}{10^{6}}$& conversion $m \to 1/eV$\\
{\small\tt Gf}& $1.66 \times 10^{-23}$&  The Fermi constant in $1/eV^2$\\
 {\small\tt Na}& $6.022 \times 10^{23}$& Avogadro number\\
{\small\tt RSun} & $6.961 \times 10^{8} \times$ m-eV & 
Radius of the Sun ( 1/eV) \\
{\small\tt REarth} & $6.378 \times 10^{6} \times$ m-eV &
 Radius of the Earth (1/eV) \\
{\small\tt Eps-Error}   &    $1.0 \times 10^{-8}$ & \\
{\small\tt Prob-Error}  &    $1.0 \times 10^{-2}$ &\\
{\small\tt N}           &   2         &  number of equations\\
{\small\tt dist }       &   0.00001   &  initial stepsize for Runge-Kutta\\
{\small\tt dist-min}    &   0.0000001 &  minimal stepsize for Runge-Kutta \\
{\small\tt EARTH}       &           0 &  Program option flag\\
{\small\tt SUN}         &  1          &  Program option flag \\
\hline 
\end{tabular}
\caption{\small
Here is a list of the most important switches and constants}
\label{}
\end{table}

\begin{table}[p]
\centering
\begin{tabular}{|l|l|}
 \hline\vspace{0.1cm}
Subroutine  & Purpose \\
[0.1cm]\hline 
{\small\tt derivs} & Computes the derivatives $dy/dx$ \\
{\small\tt runge} & Given the functions y and their derivatives $dy/dx$, it returns the advanced solution\\
{\small\tt odeint} & RK driver with adaptive stepsize control. Integrates the starting value over an interval\\
& with a required accuracy\\
{\small\tt rkqs} & Used to monitor accuracy and adjust stepsize during RK integration\\
{\small\tt rkck} & Returns advanced solution over an interval together with the estimate of truncation error\\[0.1cm]\hline 
\end{tabular}
\caption{\small
The main subroutines and functions used in the code are reported together with a brief explanation of their meaning.}
\label{}
\end{table}



\begin{thebibliography}{99}
\bibitem{revneut}
See, for instance :\\
M.~C.~Gonzalez-Garcia and Y.~Nir,
Rev.\ Mod.\ Phys.\  {\bf 75} (2003) 345;
A.~Y.~Smirnov,
arXiv:hep-ph/0306075;
S.~M.~Bilenky, C.~Giunti and W.~Grimus,
Prog.\ Part.\ Nucl.\ Phys.\  {\bf 43}, 1 (1999);
P.~Aliani, V.~Antonelli, R.~Ferrari, M.~Picariello and E.~Torrente-Lujan,
AIP Conf.\ Proc.\  {\bf 655} (2003) 103
[arXiv:hep-ph/0211062];
P.~Aliani, V.~Antonelli, R.~Ferrari, M.~Picariello and E.~Torrente-Lujan,
Les Rencontres de Physique de la Vallee d'Aoste: Results and Perspectives in 
Particle Physics. (Vol. 28,Frascati Physics Series), pp. 151-163  
[arXiv:hep-ph/0206308] ;
A.~Strumia and F.~Vissani,
Int.\ J.\ Mod.\ Phys.\ A {\bf 17} (2002) 1755
\bibitem{SNO} 
Q.~R.~Ahmad {\it et al.}  [SNO Collaboration],
Phys.\ Rev.\ Lett.\  {\bf 89} (2002) 011301 ;
Q.~R.~Ahmad {\it et al.}  [SNO Collaboration],
Phys.\ Rev.\ Lett.\  {\bf 89} (2002) 011302
\bibitem{SK} 
Y.~Koshio,
arXiv:hep-ex/0306002;
S.~Fukuda {\it et al.}  [Super-Kamiokande Collaboration],
Phys.\ Lett.\ B {\bf 539} (2002) 179

\bibitem{KamLAND} 
K.~Eguchi {\it et al.}  [KamLAND Collaboration],
Phys.\ Rev.\ Lett.\  {\bf 90} (2003) 021802
\bibitem{atmospheric} 
G.~L.~Fogli, E.~Lisi and A.~Marrone,
Nucl.\ Instrum.\ Meth.\ A {\bf 503} (2003) 179.
\bibitem{MSW} L. Wolfenstein, Phys. Rev. D {\bf 17}, 2369 (1978); 
S.P. Mikheyev and A. Yu. Smirnov, Yad. Fiz. {\bf 42}, 1441 (1985)
[Sov. H. Nucl. Phys. {\bf 42}, 913 (1985)]

\bibitem{pet3} S. T.  Petcov. {\em Phys. Lett. B} {\bf 200},3 373 (1988)
\bibitem{tov1} U. Toshev. {\em Phys. Lett. B} {\bf 196} 170 (1987)
\bibitem{2neut}
S. Parke, Phys. Rev. Lett. {\bf 57}, 1275 (1986);
P.I. Krastev and S.T. Petcov, Phys. Lett. B {\bf 207}, 64 (1988);
S.T. Petcov, Phys. Lett. B {\bf 406}, 355 (1997);
M. Bruggen, W.C. Haxton, and Y-Z Qian, Phys. Rev. D {\bf 51}, 4028 (1995);
A.B. Balantekin, J.F. Beacom, and J.M. Fetter, Phys. Lett. B {\bf 427}, 317
(1998);
G. Fiorentini, M. Lissia, G. Mezzorani, M. Moretti,
and D. Vignaud, Phys. Rev. D {\bf 49}, 6298 (1994).
\bibitem{Tor3neut}
E. Torrente Lujan, Phys. Rev. D {\bf 53}, 4030 (1996)
See also:
E.~Torrente-Lujan,
Phys.\ Rev.\ D {\bf 60} (1999) 085003;
V.~Antonelli and E.~Torrente Lujan,
Phys.\ Rev.\ A {\bf 58} (1998) 1980
\bibitem{other3neut}
T. Ohlsson and H. Snellman, hep-ph/9910546;
P. Osland and T.T. Wu, hep-ph/9912540
\bibitem{KIM}
J.S. Kim, Y.S. Chae and J.D. Kim, Comp. Phys. Comm. {\bf 120}, 41 (1999).
\bibitem{JSK}
J.S. Kim and C.W. Kim, hep-ph/9909428.
\bibitem{Kim:2000sm}
J.~S.~Kim and K.~Lee,
Comput.\ Phys.\ Commun.\  {\bf 135}, 176 (2001)
\bibitem{PDG}
K.~Hagiwara {\it et al.}  [Particle Data Group Collaboration],
Phys.\ Rev.\ D {\bf 66}, 010001 (2002)
\bibitem{bah1} J.N. Bahcall, R. Ulrich. {\em Rev. Mod. Phys.} 60 (1988) 267
\bibitem{pet2} S. T.  Petcov. {\em Phys. Lett. B} Vol.191 (1987) p.299
\bibitem{hax1} W.C. Haxton, {\em Phys. Rev. D} 35 (1987) 2352
\bibitem{not1} D. Notzold, MPI-PAE/Pth 08/87 (Munich 87)
\bibitem{ourbasic} 
P.~Aliani, V.~Antonelli, M.~Picariello and E.~Torrente-Lujan,
Nucl.\ Phys.\ B {\bf 634} (2002) 393; 
P.~Aliani, V.~Antonelli, R.~Ferrari, M.~Picariello and E.~Torrente-Lujan,
Phys.\ Rev.\ D {\bf 67} (2003) 013006;
P.~Aliani, V.~Antonelli, M.~Picariello and E.~Torrente-Lujan,
arXiv:hep-ph/0212212;
P.~Aliani, V.~Antonelli, M.~Picariello and E.~Torrente-Lujan,
JHEP {\bf 0302} (2003) 025;
P.~Aliani, V.~Antonelli, M.~Picariello and E.~Torrente-Lujan,
New J.\ Phys.\  {\bf 5} (2003) 2
\bibitem{Torsec}
E.~Torrente-Lujan,
Phys.\ Rev.\ D {\bf 59} (1999) 073001.
E.~Torrente-Lujan,
JHEP {\bf 0304} (2003) 054
E.~Torrente-Lujan,
Phys.\ Lett.\ B {\bf 389} (1996) 557.
V.~B.~Semikoz and E.~Torrente-Lujan,
Nucl.\ Phys.\ B {\bf 556} (1999) 353.
E.~Torrente-Lujan,
Phys.\ Rev.\ D {\bf 59} (1999) 093006.
\bibitem{Torthird}
S.~Khalil, C.~Munoz and E.~Torrente-Lujan;
New J.\ Phys.\  {\bf 4} (2002) 27;
D.~G.~Cerdeno, E.~Gabrielli, S.~Khalil, C.~Munoz, E.~Torrente-Lujan and E.~Torrente-Lujan,
Nucl.\ Phys.\ B {\bf 603} (2001) 231;
E.~Gabrielli, S.~Khalil and E.~Torrente-Lujan,
Nucl.\ Phys.\ B {\bf 594} (2001) 3
\bibitem{NumericalRecipes} W.~H.~Press, \emph{et al.},''Numerical Recipes in C++'', Cambridge University Press, Cambridge, 2002

\end{thebibliography}
 \end{document}